\documentclass[a4paper]{article}

\usepackage{INTERSPEECH2021}
\newcommand{\R}{\mathbb{R}}
\newcommand{\C}{\mathbb{C}}
\newcommand*{\hermconj}{^{\mathsf{H}}}
\DeclareMathOperator{\tr}{tr}

\title{Self-Attention Channel Combinator Frontend for End-to-End Multichannel Far-field Speech Recognition}
\name{Rong Gong$^1$, Carl Quillen$^2$, Dushyant Sharma$^2$, \\ Andrew Goderre$^2$, José Laínez$^3$ and Ljubomir Milanovi\'{c}$^1$}
\address{
  $^1$Nuance Communications GmbH, Vienna, Austria\\
  $^2$Nuance Communications Inc., Burlington, USA\\
  $^3$Nuance Communications S.A., Madrid, Spain}
\email{rong.gong@nuance.com}

\begin{document}

\maketitle
\begin{abstract}
When a sufficiently large far-field training data is presented, jointly optimizing a multichannel frontend and an end-to-end (E2E) Automatic Speech Recognition (ASR) backend shows promising results. Recent literature has shown traditional beamformer designs, such as MVDR (Minimum Variance Distortionless Response) or fixed beamformers can be successfully integrated as the frontend into an E2E ASR system with learnable parameters. In this work, we propose the self-attention channel combinator (SACC) ASR frontend, which leverages the self-attention mechanism to combine multichannel audio signals in the magnitude spectral domain. Experiments conducted on a multichannel playback test data shows that the SACC achieved a 9.3\% WERR compared to a state-of-the-art fixed beamformer-based frontend, both jointly optimized with a ContextNet-based ASR backend. We also demonstrate the connection between the SACC and the traditional beamformers, and analyze the intermediate outputs of the SACC.
\end{abstract}


\section{Introduction}

It has been demonstrated that multichannel ASR systems improve the recognition accuracy compared to a single channel ASR system in far-field scenarios~\cite{haeb-umbach_far-field_2021,ochiai_unified_2017,ochiai_multichannel_2017}. Existing multichannel E2E ASR systems usually comprise two parts -- the frontend and the backend. The frontend neural network takes the multichannel audio signals captured by a microphone array or distributed microphones as the input and outputs a single channel representation. The backend can be any common E2E ASR system~\cite{chan_listen_2016,hori_advances_2017,dong_speech-transformer_2018} which receives the frontend output and then produces the text tokens.

Most of the ASR frontends are based on the beamforming paradigm, which leverages the spatial information embedded in the multichannel signal to produce a denoised and dereverbed speech. Two common beamforming designs usually seen in the ASR frontend literature are MVDR-based design~\cite{ochiai_unified_2017} and non-constrained design~\cite{minhua_frequency_2019,park_robust_2020,li_neural_2016}. The former applies the MVDR formulation to calculate the beamforming coefficients, which minimizes the noise variance and imposes a constraint to not distort the signal in the beam direction~\cite{dmochowski_microphone_2010}. To calculate the MVDR beamformer, one needs to know two covariance matrices formulated by the clean speech and the background noise. However, the signal captured by the microphones is a mixture of the reverberant speech and the noise, so the separated clean speech and noise are not given by default. Thus, various studies explored the methods for estimating a better covariance matrix, e.g. designing features and neural networks to predict the speech and noise masks~\cite{ochiai_unified_2017,ochiai_multichannel_2017,zhang_end--end_2020}, improving the numerical stability when calculating the inverse of the noise covariance matrix~\cite{zhang_end--end_2021}. Conversely, the non-constrained design doesn't impose any constraint when estimating the beamforming coefficients, which does so by e.g. using dense neural networks to learn fixed beamformers in the frequency domain for multiple beam directions, then choosing one of the directions or combining them~\cite{minhua_frequency_2019, park_robust_2020}; using recurrent neural networks (RNNs) to learn adaptive beamformers in the time domain~\cite{li_neural_2016}.

Another ASR frontend research stream leverages various attention mechanisms which generate the weights to either select or combine the multichannel signal. These methods do not utilize the beamforming paradigm, and are closely relevant to our work. S. Kim and I. Lane~\cite{kim_end--end_2017} make use of a custom designed RNN to combine the multichannel Mel filterbank features. S. Braun, et al.~\cite{braun_multi-channel_2018} use a long short-term memory (LSTM) network to combine the multichannel spectrograms. T. Ochiai et al.~\cite{ochiai_unified_2017} utilize a dense network for reference microphone selection. 

Self-attention mechanism has been previously applied to multichannel speech enhancement~\cite{tolooshams_channel-attention_2020} and recently to mulichannel E2E ASR~\cite{chang_end--end_2021}. The latter modified the ASR encoder and decoder to be able to process the multichannel signal in the backend. Thus, it requires a large amount of multichannel audio data for model training, which is not always available. This paper proposes a simple self-attention based ASR frontend for multichannel signal combination. This frontend can be jointly optimized with any common ASR backend which is either randomly initialized or pretrained with single channel data.

The rest of the paper is organized as follows: Section 2 introduces the proposed frontend -- SACC and discusses its connection to beamforming. Section 3 describes the dataset, experimental baselines, setup, and evaluation metric. Section 4 demonstrates the results of the baselines and the SACC. Section 5 analyzes the intermediate outputs of the SACC frontend. Finally, Section 6 concludes the paper and discusses future directions.

\section{Self-attention channel combinator}

\subsection{Problem description}
Let \(\mathbf{x} \in \R^{N \times \C} \) be the discrete-time \(N\) samples signal captured by multiple microphones with the channel number \(C\). Let \(\mathbf{X} = [ \mathbf{X}^1, \dots, \mathbf{X}^C ] \in \C^{T \times C \times F}\) be the multichannel short-time Fourier transform (STFT) and \(\mathbf{X}^{mag} \in \R^{T \times C \times F}\) be the magnitude, where \(T, F\) are the time frames and the number of frequency bins. 

The channel combinator estimates weights \(\mathbf{w} \in \R^{T \times C \times 1}\) to produce a single channel representation \(\mathbf{S} \in \R^{T \times F}\), which is done by the element-wise multiplication of the weights \(\mathbf{w}\) and \(\mathbf{X}^{mag}\), and then the sum over the channel dimension. The subsequent sections will describe the SACC network architecture and how it is jointly optimized with the ASR backend.

\subsection{Network architecture} \label{sec:na}

\begin{figure}[t]
  \centering
  \includegraphics[width=\linewidth]{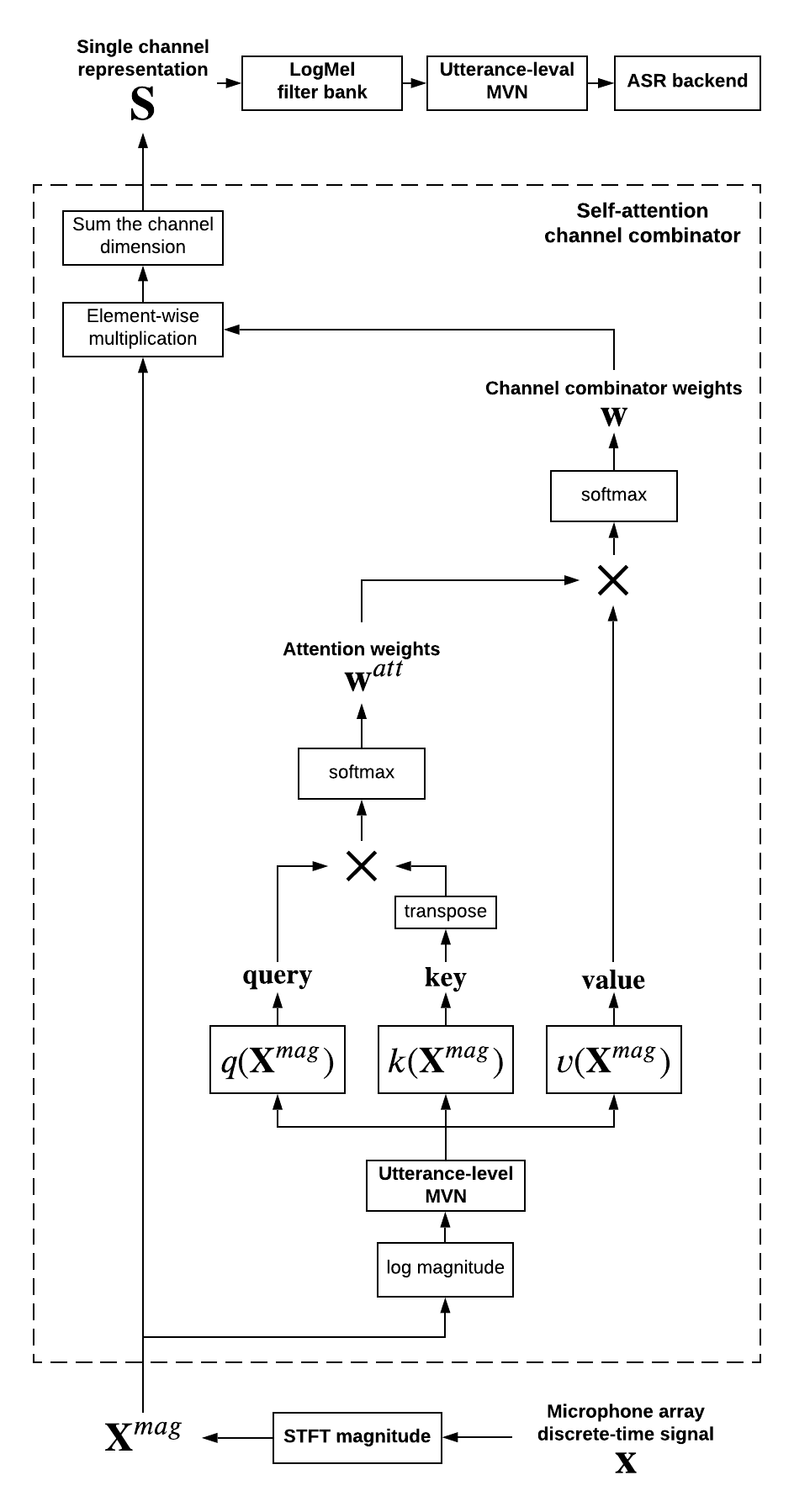}
  \caption{Self-attention channel combinator flowchart.}
  \label{fig:sa-flowchart}
\end{figure}

A flowchart of the SACC network architecture is given in Figure~\ref{fig:sa-flowchart}. A self-attention mechanism is applied to the STFT magnitude \(\mathbf{X}^{mag}\) to produce the channel combinator weights \(\mathbf{w}\).

To compute the \(\mathbf{query}, \mathbf{key}\) and \(\mathbf{value}\) tensors for the self-attention mechanism, we convert \(\mathbf{X}^{mag}\) into the logarithmic scale and then perform the utterance-level mean and variance normalization (MVN) on each frequency bin. 
\(q, k, v\) are three dense layers with linear activations. They transform the normalized log magnitude \(\mathbf{X}^{mag}\) into \(\mathbf{query} \in \R^{T \times C \times D}, \mathbf{key} \in \R^{T \times C \times D}\) and \(\mathbf{value} \in \R^{T \times C \times 1}\), where \(D\) is the dimension of the linear transform of \(q\) and \(k\), aka the units of the dense layer. As we would like to have channel combinator weights \(\mathbf{w}\) that work homogeneously on all frequency bins, we use the dense layer \(v\) with a single unit to contract the frequency dimension of \(\mathbf{X}^{mag}\) and to produce the \(\mathbf{value}\).

The self-attention weights \(\mathbf{w}^{att} \in \R^{T \times C \times C} \) is calculated by Eq.~\ref{eq:sa}, where the softmax is applied on the last channel dimension of the tensor product. The element \(\mathbf{w}^{att}_{tij}\) can be seen as the cosine similarity between the channel \(i\) of the query -- \(\mathbf{query}_i\) and the channel \(j\) of the key -- \(\mathbf{key}_j\) at the time frame \(t\), assuming that both \(\mathbf{query}_i\) and \(\mathbf{key}_j\) are normalized.

\begin{equation}
  \mathbf{w}^{att} = \mathrm{softmax}(\frac{\mathbf{query} (\mathbf{key})^\intercal}{\sqrt{D}})
  \label{eq:sa}
\end{equation}

The channel combinator weights is calculated by Eq.~\ref{eq:ccw}, where the softmax is applied on the channel dimension of the tensor product between \(\mathbf{w}^{att}\) and \(\mathbf{value}\).

\begin{equation}
  \mathbf{w} = \mathrm{softmax}(\mathbf{w}^{att} \mathbf{value})
  \label{eq:ccw}
\end{equation}

The single channel representation \(\mathbf{S}\) is calculated by Eq.~\ref{eq:sum}. To achieve the element-wise multiplication between \(\mathbf{w}\) and \(\mathbf{X}^{mag}\), the last dimension of \(\mathbf{w}\) is broadcasted to the number of the frequency bins in \(\mathbf{X}^{mag}\). Finally, we sum over the channel dimension to produce the single channel representation.

\begin{equation}
  \mathbf{S} = \sum_{c} \mathbf{w} \odot \mathbf{X}^{mag}
  \label{eq:sum}
\end{equation}
where \(\odot\) indicates element-wise multiplication.

\(\mathbf{S}\) is a weighted sum over the channels of \(\mathbf{X}^{mag}\) at each time frame. We calculate its logarithmic Mel (LogMel) representation and then conduct the utterance-level MVN, which results in the input feature for the ASR backend. Consequently, the SACC and the ASR backend can be jointly optimized.

\subsection{Connection to beamforming}

Beamforming is a type of channel combinator. One of the goals for the beamforming is to combine multichannel microphone signal into a single channel signal. Let \(\mathbf{h} = [\mathbf{h}^1, \dots, \mathbf{h}^C] \in \C^{F \times C}\) be the beamformer weights and \(\mathbf{h}_f \in \C^{1 \times C}\) be the weights at the frequency bin \(f\). The application of the beamformer on \(\mathbf{X}\) leads to~\cite{dmochowski_microphone_2010}:
\begin{equation}
  \mathbf{Y}_f = \mathbf{X}_f \mathbf{h}_f\hermconj
  \label{eq:beamforming}
\end{equation}
where \(\mathbf{X}_f\) is the component of \(\mathbf{X}\) at the frequency bin \(f\) and \(\mathbf{Y}_f\ \in \C^{T})\) is the beamformed spectrogram. \(\mathsf{H}\) indicates Hermitian conjugate. 

Eq.~\ref{eq:beamforming} shows that \(\mathbf{Y}_f\) is the weighted sum of the channels in  \(\mathbf{X}_f\), and the weights are \(\mathbf{h}_f\). Since both \(\mathbf{Y}_f\) and \(\mathbf{h}_f\) are of complex values, a beamformer not only applies the weights on the magnitude but also shifts the phase of the \(\mathbf{X}_f\), whereas the SACC only applies the weights on the magnitude.

Although the beamformer weights can be estimated on the time frame-level~\cite{higuchi_frame-by-frame_2018} or the segment-level~\cite{boeddeker_exploring_2018}, when jointly optimizing them with the ASR backend, due to the high computational complexity, the weights are usually calculated on the utterance-level. The SACC estimates the weights on the time frame-level.

From the mathematical perspective, the SACC resembles the MVDR beamformer formulation. The MVDR method estimates the beamformer coefficients by solving a constrained minimization problem and results in the following formula~\cite{dmochowski_microphone_2010}:
\begin{equation}
  \mathbf{h}_f = \frac{\mathbf{\Phi}_v^{-1} \mathbf{\Phi}_s \mathbf{u}}{\tr [\mathbf{\Phi}_v^{-1} \mathbf{\Phi}_s]}
  \label{eq:mvdr}
\end{equation}
where the \(\mathbf{\Phi}_v , \mathbf{\Phi}_s \in \C^{C \times C} \) are respectively the noise and speech covariance matrices with the subscript \(f\) omitted. \(\mathbf{u} \in \R^{C \times 1}\) is the one-hot vector for the reference microphone selection. The term \(\mathbf{\Phi}_s \mathbf{u} \in \C^{C \times 1}\) can be seen as the acoustic transfer function relative to the reference microphone channel. The denominator is a normalization scalar. In the SACC, the dimensionality of the self-attention weights \(\mathbf{w}^{att} \in \R^{T \times C \times C}\) resembles that of \(\mathbf{\Phi}_v^{-1}\), and the dimensionality of \(\mathbf{value} \in \R^{T \times C \times 1}\) resembles that of \(\mathbf{\Phi}_s \mathbf{u}\). 

Finally, Eqs.~\ref{eq:beamforming} and~\ref{eq:mvdr} demonstrate the narrow-band beamformer~\cite{dmochowski_microphone_2010}, where the weights are estimated independently on each frequency bins. On the contrary, the SACC applies a set of weights homogeneously on all frequency bins. 

\section{Experiments}

In the following we describe the training and test data used in this work along with the baseline algorithms and the experimental setup.

\subsection{Training data}

The training data is based on a 460~hours subset of speech from the clean training partitions of Librispeech~\cite{Panayotov2015-LAA} and the English partition of Mozilla Common Voice\footnote{https://commonvoice.mozilla.org/en}. The subset was selected based on various signal characteristics extracted using the NISA~\cite{Sharma2020-NIE} algorithm and placing thresholds on estimated C50, SNR and PESQ scores\footnote{25~dB for C50 and SNR; 3.0 for PESQ} to ensure the base material had minimal reverberation, noise and and high perceptual quality. This base material was convolved with room impulse responses (RIRs) generated using the Image method~\cite{Allen1979-IMF}. A number of room configurations representing typical meeting room dimensions with T60 in the range {[}0.27 to 0.79~s{]} were simulated. Within each room, an 8 channel Uniform Linear Array (ULA) with a 33~mm inter-mic.\ spacing and an omni-directional source were placed in random positions, resulting in a total of 40,000 RIRs. Each utterance from the source data was convolved with a randomly selected RIR, followed by addition of ambient, babble and fan noise {(}with an SNR sampled uniformly from 3 to 25~dB{)}. Microphone self noise was simulated by adding white noise at a 45~dB SNR to each microphone channel followed by a random gain offset in the range 0.1 to 2.0~dB was applied to each microphone of the ULA to simulate inter-mic.\ gain variations. Finally, an overall level augmentation was applied to all utterances in the {[}-1 to -15~dBFS {]} range, to make the training process robust to level variations in the data.

\subsection{Test data}
In order to make the evaluations realistic, simultaneous playback and recording of speech material was collected in a typical meeting room. The room was setup with an analogue eight channels ULA with 33~mm inter mic.\ spacing, mounted on a wall and four artificial mouth loudspeakers were placed in four positions. The loudspeaker playback signals were recorded at 48~kHz sample rate, and the recordings were then down-sampled to 16~kHz. A subset of the Librispeech~\cite{Panayotov2015-LAA} clean test partition was used as the playback data {(}to ensure only clean utterances were played back, the same selection criteria was applied as used in training data{)}. The testing partition has no overlap with training data in terms of utterances or speakers. The same set of utterances were played back individually in four positions to form four sub-testsets.

\subsection{Baselines}
We compare the SACC with four baselines: (1) single channel distant microphone (SDM): We choose the middle channel (the 4th microphone) signal in the datasets, and feed it into the ASR backend for training and testing. (2) Random channel distant microphone (RDM): For each utterance in the training dataset, we uniform-randomly choose a channel for ASR backend training, and always choose the middle channel for testing. (3) MVDR beamformer (MVDR): This beamformer utilizes the MVDR formulation~\cite{dmochowski_microphone_2010} and takes all channels as the input. For a better estimation of the speech covariance matrix in a diffuse noise field, the beamformer input is filtered by a coherence-to-diffuse ratio (CDR) based mask~\cite{schwarz_coherent--diffuse_2015}. The MVDR beamformer is applied as a data preprocessing step for both ASR training and testing. (4) Neural beamforming (NBF)~\cite{minhua_frequency_2019}: This baseline learns fixed beamformer weights in eight beam directions, which are utilized to filtering the array signal. The filtered signal energy in eight directions are then combined to form a single channel representation. The NBF takes all array channels as the input and is jointly optimized with the ASR backend.

\subsection{Experimental setup and Evaluation}

The ASR backend is an attention-based encoder-decoder (AED) system. The encoder is a variant of the one described in the ContextNet paper~\cite{han_contextnet_2020}, and the decoder is a single layer LSTM network. SpecAugment~\cite{park_specaugment_2019} is applied on the utterances before feeding them into the ASR training. The STFT window size and hop size for all the baselines and the SACC are 25ms and 10ms. The SACC has \(D=256\). The LogMel utilizes 64 filter banks that spans from 0 to 8kHz.

The ASR uses 5k subword tokens for the transcriptions. The Adam optimizer and Noam learning rate scheduler~\cite{vaswani_attention_2017} is applied during the training, and the label-smoothed cross-entropy loss~\cite{chorowski_towards_2017} is minimized between the ground truth tokens and the predictions. For all experiments, we train the system for a maximum of 70 epochs with the early stopping patience of 5 epochs.

Results of the experiments are evaluated by the word error rate (WER) and the WER reduction (WERR).

\section{Results}

Table~\ref{tab:results} shows the results of the experiments, where the WERs are the averaged values over four speaker positions in the test set. The first interesting observation is that the RDM performs on par with the MVDR. The RDM can be seen as a regularization for the ASR backend training, which provides to the ASR more data variability and results in a more robust ASR model than the SDM training. This variability might include small differences in the acoustic transfer function from a source to each microphone. The MVDR is a common beamformer design, which is able to produce an enhanced speech with a better quality. This observation indicates that having a robust ASR backend is as important as a well-designed ASR frontend.

\begin{table}[th]
    \caption{The four speaker positions averaged WERs (\%) and the WERRs (\%) by comparing to the SDM. \#channels: the number of the microphone array channels utilized for the frontends. \#params: the number of parameters of the frontends.}
    \label{tab:results}
    \centering
    \scalebox{0.9}{
    \begin{tabular}{lcccc}
        \toprule
        Methods & \#channels & \#params & WER  & WERR \\
        \midrule
        SDM     & 1          & -        & 11.9 & -    \\
        RDM     & 1          & -        & 10.9 & 8.4  \\
        MVDR    & 8          & -        & 11.0 & 8.3  \\
        NBF     & 8          & 90.5k    & 10.3 & 13.4 \\
        SACC    & 8          & 132.4k   & 9.2  & 22.7 \\
        \bottomrule
    \end{tabular}
    }
\end{table}

The second observation is that the NBF and SACC, both contain small numbers of parameters, outperform the MVDR. The NBF and SACC are jointly optimized with the ASR backend, but the MVDR is not because it doesn't contain learnable parameters. This indicates that, in our experimental context, the joint optimization of the ASR frontend and backend is preferable to an ASR trained with the MVDR beamformed data.

Lastly, we observe that the SACC achieves a 9.3\% WERR increase than the NBF, which suggests its improved effectiveness as an ASR frontend.

\section{Analysis}

In this section, we take an example from the test data to demonstrate the intermediate outputs of the SACC frontend.

\begin{figure}[th]
  \centering
  \includegraphics[width=0.95\linewidth]{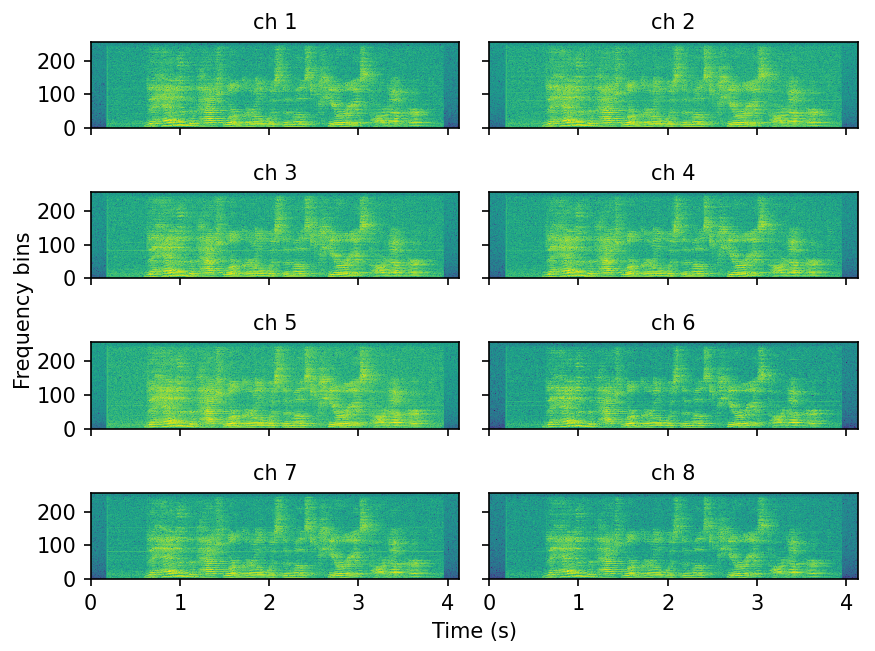}
  \caption{Utterance-level normalized eights channels log-scale magnitude spectrogram. ch: channel. }
  \label{fig:a0}
\end{figure}

Figure~\ref{fig:a0} shows the log-scale \(\mathbf{X}^{mag}\) after utterance-level MVN for this example. It can be seen that different channels have different spectrogram characteristics, e.g. the channel 5 exhibits higher speech energy than the channel 8.

\begin{figure}[th]
  \centering
  \includegraphics[width=0.9\linewidth]{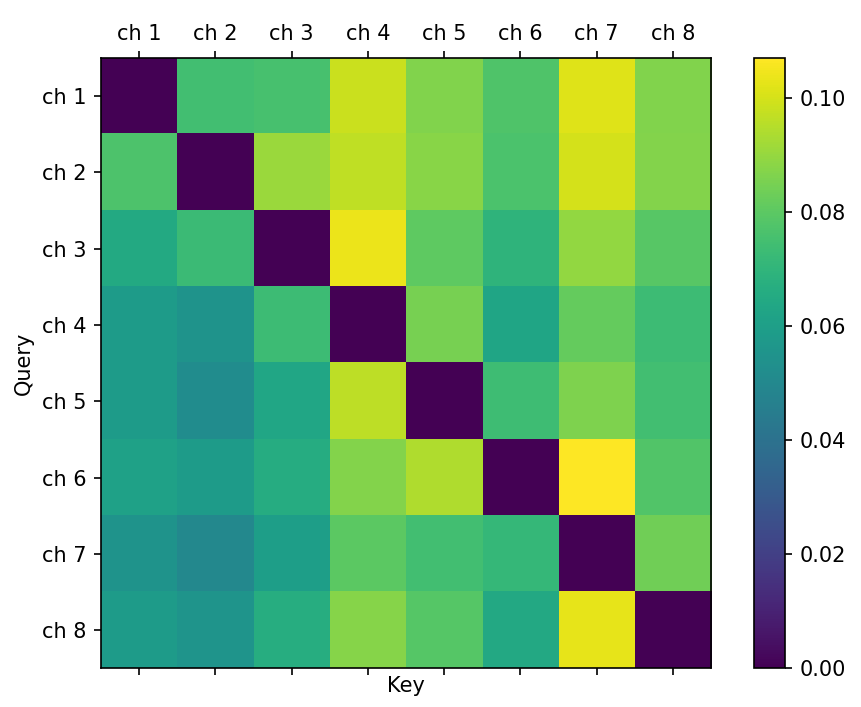}
  \caption{Time-averaged self-attention weights \(\mathbf{w}^{att}\). For a better visualization, the diagonal entries are set to 0. ch: channel.}
  \label{fig:a1}
\end{figure}

Figure~\ref{fig:a1} demonstrates the time-averaged attention weights \(\mathbf{w}^{att}\) for this example. As mentioned in section~\ref{sec:na}, the weights element \(\mathbf{w}^{att}_{tij}\) can be seen as the cosine similarity between \(\mathbf{query}_i\) and \(\mathbf{key}_j\) at the time frame \(t\). Thus, Figure~\ref{fig:a1} represents the averaged similarity matrix over the frames of this example. Since the large values of the matrix diagonal entries overshadow the off-diagonal ones, for a better visualization, we set the diagonal entries to 0.

We observe clearly that some entries have larger values than the others when comparing in the same row, e.g. \(\mathbf{w}^{att}_{3,4}\), \(\mathbf{w}^{att}_{6,7}\) and \(\mathbf{w}^{att}_{8,7}\), which indicates the query and key pairs represented by these entries are more similar than the other pairs. Consequently, when producing the channel combinator weights \(\mathbf{w}\) at the corresponding channel, the values of these entries account for more weights than the others in the same row. E.g. when producing \(\mathbf{w}\) at the channel 8, the \(\mathbf{value}_7\) and \(\mathbf{value}_8\) account for more weights than the other values.

\begin{figure}[th]
  \centering
  \includegraphics[width=\linewidth]{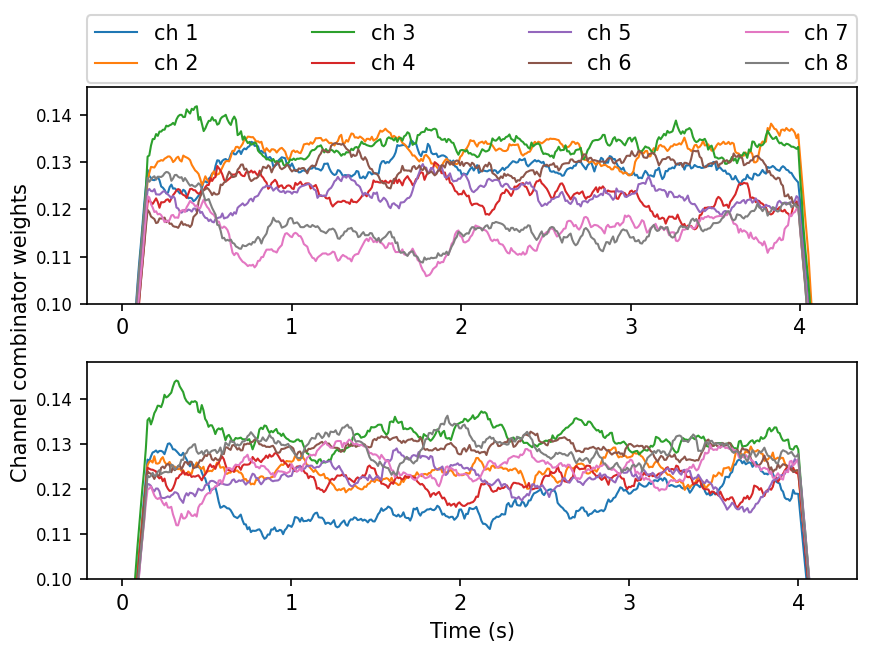}
  \caption{Channel combinator weights \(\mathbf{w}\) per channel. Upper: the same example in Figure~\ref{fig:a0} and ~\ref{fig:a1}. Bottom: this example is the same utterance as in the upper plot, but recorded with the loudspeaker played back in a different position. ch: channel.}
  \label{fig:a2}
\end{figure}

The upper plot in Figure~\ref{fig:a2} shows the combinator weights \(\mathbf{w}\) per channel for the same example in Figure~\ref{fig:a0} and~\ref{fig:a1}. For a better visualization, we applied on \(\mathbf{w}\) a moving average with a size of 30 time frames. We observe that \(\mathbf{w}\) fluctuates within the range of [0.10 - 0.14]. Overall, when producing \(\mathbf{S}\), the channels 2 and 3 of \(\mathbf{X}^{mag}\) account for more weights, while the channels 7 and 8 account for less weights than the other channels. The lower plot in Figure~\ref{fig:a2} shows an example of the same utterance as in the upper figure, but recorded with the loudspeaker played back in a different position. We observe that the weights are position-varying, and the channels 7 and 8 no longer account for less weights than the other channels.

\section{Conclusion}

We proposed the SACC frontend for E2E multichannel ASR, and demonstrated that the frontend can estimate channel combination weights via the self-attention mechanism. The experiments showed that the SACC outperformed a traditional MVDR beamformer and a state-of-the-art neural beamforming frontend in terms of the WER.

\bibliographystyle{IEEEtran}

\bibliography{mybib}

\end{document}